\documentclass[aps,prb,reprint,amssymb,amsmath,
superscriptaddress,longbibliography]{revtex4-2}

\usepackage{lipsum, xcolor,graphicx, amsmath}
\usepackage{placeins, float}
\usepackage{amssymb}    
\usepackage{comment}
\usepackage{braket}
\usepackage{balance}
\usepackage{lineno}

\newcommand{\red}[1]{\textcolor{black}{#1}}
\begin{document}


\title{Direct Evidence of a Near-Ideal $J_\mathrm{eff}=1/2$ Ground State in Triangular-Lattice Na$_2$BaCo(PO$_4$)$_2$}

\author{M. M. Ferreira-Carvalho}

\affiliation{Max Planck Institute for Chemical Physics of Solids, N{\"o}thnitzer Str. 40, 01187 Dresden, Germany}
\affiliation{Institute of Physics II, University of Cologne, Zülpicher Straße 77, 50937 Cologne, Germany }

\author{S.H. Chen}

\affiliation{Max Planck Institute for Chemical Physics of Solids, N{\"o}thnitzer Str. 40, 01187 Dresden, Germany}
\author{Y.~C.~Ku}

\affiliation{Department of Electrophysics, National Yang Ming Chiao Tung University, Hsinchu, Taiwan}
\affiliation{National Synchrotron Radiation Research Center,
Hsinchu, Taiwan}
 \author{Anagha Jose}
 \affiliation{Institut f{\"u}r Festk{\"o}rperphysik, Leibniz IFW Dresden, D-01069 Dresden, Germany}
 \author{Ryan Morrow}
 \affiliation{Institut f{\"u}r Festk{\"o}rperphysik, Leibniz IFW Dresden, D-01069 Dresden, Germany}

\author{C.~Y.~Kuo}

\affiliation{Department of Electrophysics, National Yang Ming Chiao Tung University, Hsinchu, Taiwan}
\affiliation{National Synchrotron Radiation Research Center,
Hsinchu, Taiwan}

\author{C.~F.~Chang}

\affiliation{Max Planck Institute for Chemical Physics of Solids, N{\"o}thnitzer Str. 40, 01187 Dresden, Germany}

\author{Z. Hu}

\affiliation{Max Planck Institute for Chemical Physics of Solids, N{\"o}thnitzer Str. 40, 01187 Dresden, Germany}

\author{M. W. Haverkort}
\affiliation{Institute for theoretical physics, Heidelberg University, Philosophenweg 19, 69120 Heidelberg, Germany}

\author{L. H. Tjeng}

\affiliation{Max Planck Institute for Chemical Physics of Solids, N{\"o}thnitzer Str. 40, 01187 Dresden, Germany}

\date{\today}

\begin{abstract}

We investigated the local Co $3d$ electronic structure of Na$_2$BaCo(PO$_4$)$_2$ using polarization-dependent X-ray absorption spectroscopy (XAS) in combination with full multiplet cluster calculations. We employed the line-fitting inverse partial fluorescence yield (IPFY) technique to obtain accurate XAS spectra from strong insulating materials. Our combined experimental and theoretical analysis reveals a very small effective trigonal distortion of only 11 meV in the CoO$_6$ octahedra, indicating a close to ideal condition to render a ground state with the $J_\mathrm{eff}=1/2$ character. With our cluster model we were also able to simulate magnetic susceptibility measurements along different directions in the crystal. These findings highlight Na$_2$BaCo(PO$_4$)$_2$ as a promising platform for exploring exotic magnetic phenomena associated with $J_\mathrm{eff}=1/2$ ground states on triangular lattices.

\end{abstract}


\maketitle

\section{Introduction}
Frustrated quantum magnets provide a rich platform for exploring unconventional quantum phases driven by competing interactions and reduced dimensionality. Among these, two-dimensional $S=1/2$ triangular lattices have been the focus of many theoretical and experimental studies of frustration-induced phenomena \cite{ANDERSON1973153,chubukovQuantumTheoryAntiferromagnet1991,starykhUnusualOrderedPhases2015}. In such systems, geometric frustration and quantum fluctuations act together to destabilize conventional magnetic order, enabling the possibility of exotic ground states. While early theoretical work showed that the  $S=1/2$  Heisenberg antiferromagnets on a triangular lattice exhibits a 120° non-collinear ordered state \cite{bernuSignatureNeelOrder1992,bernuNeelOrderSpin1993,bernuExactSpectraSpin1994,whiteNeelOrderSquare2007}, later studies demonstrated that long-range order can be suppressed through extended exchange interactions, with emerging anisotropic interactions \cite{bauerSchwingerbosonMeanfieldStudy2017,ferrariDynamicalStructureFactor2019,gongGlobalPhaseDiagram2017,huDiracSpinLiquid2019,iqbalSpinLiquidNature2016,maksimovAnisotropicExchangeMagnetsTriangular2019,saadatmandDetectionCharacterizationSymmetrybroken2017,zhuSpinLiquidPhase2015,zhuTopographySpinLiquids2018}.

Recently, triangular-lattice cobaltates with Co$^{2+}$ ions in octahedral coordination have attracted strong interest due to their interplay of frustration and spin orbit coupling (SOC) \cite{liSpinLiquidsGeometrically2020}. In these $d^7$  systems, crystal field effects and SOC can stabilize a well-isolated, Kramers-degenerate $J_{\mathrm{eff}}$ =1/2 ground state. This pseudospin state in a triangular lattice creates an ideal environment for studying potential emergent quantum phases.

\begin{figure}
	\centering
	\includegraphics[width=\linewidth]{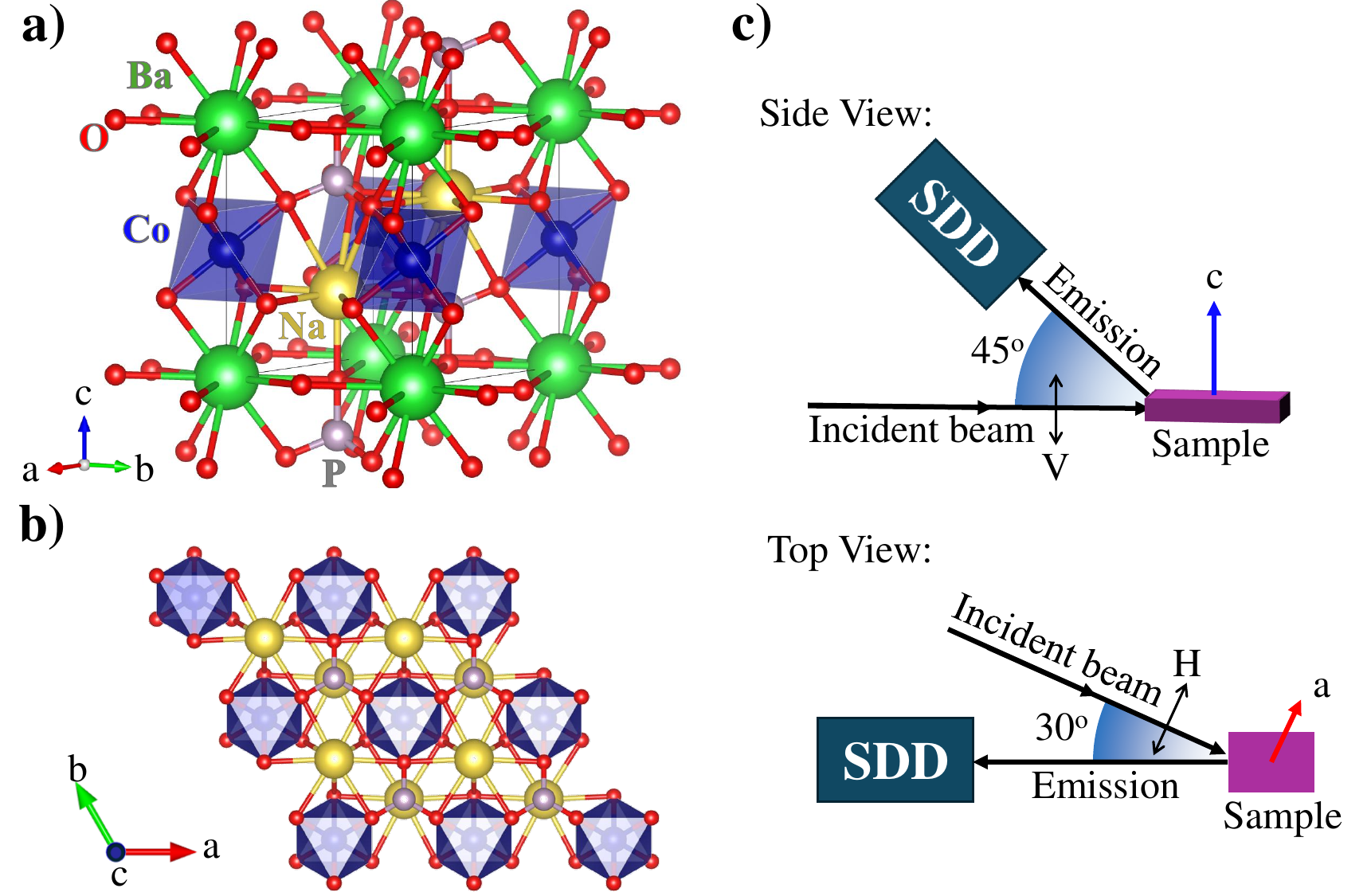}
	\caption{(a) Crystal structure of Na$_2$BaCo(PO$_4$)$_2$, showing layers of CoO$_6$ octahedra stacked along the \textit{c}-axis. (b) The \textit{ab} plane containing the Co$^{2+}$ triangular lattice. (c) Schematic of the experimental geometry, illustrating the orientation of the incident X-ray beam, linear polarization directions, and the silicon drift detector (SDD) relative to the sample.
	}
	\label{fig:experimental setup}
\end{figure}

In this study we investigate Na$_2$BaCo(PO$_4$)$_2$. Figure~\ref{fig:experimental setup} (a) and (b) display the $Pm\bar{3}1$ structure crystal structure and the triangular lattice arrangement of the Co ions in the $ab$ plane. This compound has been recently the subject of intense and conflicting research, reflecting the complexity of its magnetic ground state. Early debates centered on the presence or absence of long-range magnetic order, with several studies supporting the emergence of a quantum spin liquid (QSL) phase. These include an initially reported exceptionally low ordering temperature of approximately 50 mK \cite{zhongStrongQuantumFluctuations2019}, later revised to around 148 mK alongside evidence of a nonzero residual thermal conductivity \cite{liPossibleItinerantExcitations2020}, as well as the observation of spinon-like excitations \cite{leeTemporalFieldEvolution2021}.

However, other studies have presented less compelling evidence for a QSL, pointing instead to more conventional magnetic behavior \cite{huangThermalConductivityTriangularlattice2022,woodlandContinuumExcitationsSharp2025,shengContinuumSpinExcitations2025}. More recently, the focus has shifted to consider this material as a promising candidate for the realization of the long-sought spin-supersolid state \cite{gaoSpinSupersolidityNearly2022,xiangGiantMagnetocaloricEffect2024}.

The core of this ongoing debate lies in the details of the interplay between geometric frustration and strong SOC. The delicate balance is highly sensitive to small changes in local structure or interactions. To clarify the microscopic origin of unconventional magnetism and to assess its suitability as a Kitaev/spin-liquid system, a direct local probe of the local Co$^{2+}$ electronic structure is necessary. It is known that Co $d^7$ in octahedral coordination displays low lying excited states that are easily populated with increasing temperature, rendering Curie-Weiss fit of magnetic susceptibility not reliable for the determination of ground state properties \cite{burnusLocalElectronicStructure2008}. It is also important to note that a distortion of the CoO$_6$ octahedra has a strong impact on the electronic properties. \red{There is however confusion in the literature about the sign of the trigonal distortion, i.e. whether the $t_{2g}$ hole is more in the $e_{g}^\pi$ or in the $a_{1g}$ orbital \cite{wellmFrustrationEnhancedKitaev2021,mouComparativeRamanScattering2024,hussainExperimentalEvidenceCrystalfield2025a,popescuZeemanSplitKramers2025}.}

Here, we report on our experimental study on Na$_2$BaCo(PO$_4$)$_2$ single crystals using polarization dependent soft X-ray absorption spectroscopy (XAS) at the Co $L_{2,3}$ edge, \red{an element specific} technique which is inherently sensitive to the local Co $3d$ electronic structure. Importantly, dipole selection rules govern the accessibility and intensity of distinct final states, making XAS a powerful probe of the ground state and low-lying excited states when combined with theoretical calculations. \red{XAS is a high-energy spectroscopic technique and can determine the interaction parameters like crystal fields, covalency, SOC, and local distortions that build up the low energy excitations} \cite{tanakaResonant3d3p1994,degrootXrayAbsorptionDichroism1994,burnusValenceSpinOrbital2006,burnusLocalElectronicStructure2008,linLocalOrbitalOccupation2010,chinSpinorbitCouplingCrystalfield2019}. For our Co$^{2+}$ case, the technique is extremely sensitive to which of the lowest $J_{\text{eff}} = \frac{1}{2}, \frac{3}{2},$ or $\frac{5}{2}$ states are occupied \red{and by using the temperature dependence it provides access to the excited states}. The polarization dependence is in particular powerful to shed light on the symmetry and magnitude of the non-octahedral local distortion and the resulting $3d$ orbital occupation. 

Such soft XAS measurements can be challenging, owing to the fact that insulating samples prove to be non-ideal for acquisition of reliable data due to charging effects in the commonly used total electron yield mode, or due to self absorption effects in the total fluorescence yield mode. In order to overcome such limitations but still make use of the strengths of the technique, we employed inverse partial fluorescence yield method \cite{achkarBulkSensitiveXray2011,achkarDeterminationTotalXray2011}, which is bulk sensitive and not prone to charging or self absorption effects. To obtain the desired information concerning the local Co $d$ quantum numbers, we performed a theoretical analysis of the measured spectra using configuration interaction cluster calculations. Furthermore, the CI-derived ground state, low-lying excited states, and their energy splittings reproduce the measured temperature-dependent susceptibility, confirming the internal consistency of these parameters.

\section{Methods}

Single crystals of Na$_2$BaCo(PO$_4$)$_2$ were prepared from stoichiometric amounts of (NH$_4$)$_2$HPO$_4$, BaCO$_3$, Na$_2$CO$_3$, and CoCO$_3$, with additional NaCl added as a flux. The mixture was thoroughly ground and heat-treated in an Al$_2$O$_3$ crucible using a box furnace. It was heated to 950\textdegree C and then slowly cooled to 750\textdegree C at a rate of 3\textdegree C/h. Pinkish, transparent thin-plate crystals were obtained with typical dimensions of 2x3 mm$^2$ in-plane and $\sim0.1$ mm thickness, consistent with a highly insulating, layered material.

The XAS experiment was conducted at beamline TPS~45A1 of the National Synchrotron Radiation Research Center in Taiwan~\cite{tsaiSubmicronSoftXray2019a}. The experimental geometry is illustrated in Figure~\ref{fig:experimental setup} (c). The Na$_2$BaCo(PO$_4$)$_2$ sample was mounted in an aluminum sample holder and cleaved in a preparation chamber with a base pressure of 10$^{-10}$~mbar to obtain a clean surface. It was then transferred in vacuo to the measurement chamber, which had a base pressure of 10$^{-11}$~mbar.

During the measurement, the sample was oriented such that the linearly horizontally polarized soft X-rays (LH) polarization was aligned parallel to the $a$-axis, while the linearly vertically polarized X-rays (LV) polarization was parallel to the $c$-axis. The beam spot size in the vertical direction was set to 7 $\mu$m to accommodate the platelet thickness of $\sim$ 100 $\mu$m. A CoO reference sample was measured simultaneously upstream of the beam to allow for precise energy calibration. For each scan the incident beam polarization was set to either LH or LV by rotating the polarization of the X-ray beam, allowing the measurement of the linear dichroism (LD) between the in-plane and out of plane directions without changing the beam spot on the sample. The spectra were collected at \textit{T}= 300 K, 200 K, 100 K and 40 K.

A further improvement of the iPFY technique, namely by line fitting the emission spectra as described in more detail in~\cite{ferreira-carvalhoTrigonalDistortionKitaev2025}, was used to accurately capture the spectral line shape and relative peak intensities. This improved procedure is necessary in situations where the incident beam energy encompasses absorption edges of elements with emission energies close to the monitored O-$K$ emission, in the present case, Ba $M_\alpha$. 

For the simulation of the XAS spectra and magnetic susceptibility, \red{we carried out the well-proven full-multiplet configuration-interaction calculations using a CoO$_6$ cluster. It accounts for the intra-atomic Co $3d$-$3d$ and $2p$-$3d$ Coulomb interactions, the atomic Co $2p$ and $3d$ spin-orbit couplings, the O $2p$ - Co $3d$ hybridization, and the local crystal ﬁeld \cite{tanakaResonant3d3p1994, degrootXrayAbsorptionDichroism1994}}. We used the \textsc{Quanty} code~\cite{haverkortMultipletLigandfieldTheory2012, haverkortBandsResonancesEdge2014, luEfficientRealfrequencySolver2014,Quanty}. \red{Parameters for the multipole part of the Coulomb interactions and the SOC were given by the Hartree-Fock values \cite{cowanTheoryAtomicStructure1981}, while the monopole parts as well as the O 2$p$ to Co 3$d$ charge transfer energies were estimated from photoemission experiments on typical Co$^{2+}$ compounds \cite{bocquetElectronicStructureEarly1996}. The SOC value of 66 meV for the Co $3d^7$ configuration was shown to reproduce well the XAS spectra and ground state properties of Co$^{2+}$ oxides \cite{csiszarControllingOrbitalMoment2005b, burnusLocalElectronicStructure2008, hollmannLocalSymmetryMagnetic2010}}.

The hybridization parameters for the full-multiplet configuration-interaction cluster calculations were derived from a tight-binding model constructed from an \textit{ab initio} Wannierized LDA band structure, obtained using the FPLO code~\cite{koepernikFullpotentialNonorthogonalLocalorbital1999, fplo21}, and then scaled to fit the experimental data.

Magnetic susceptibility measurements were performed using a Quantum Design MPMS-XL (model CXL398CE) SQUID magnetometer.

\section{X-ray absorption}
\begin{figure}
    \centering
    \includegraphics[width=\linewidth]{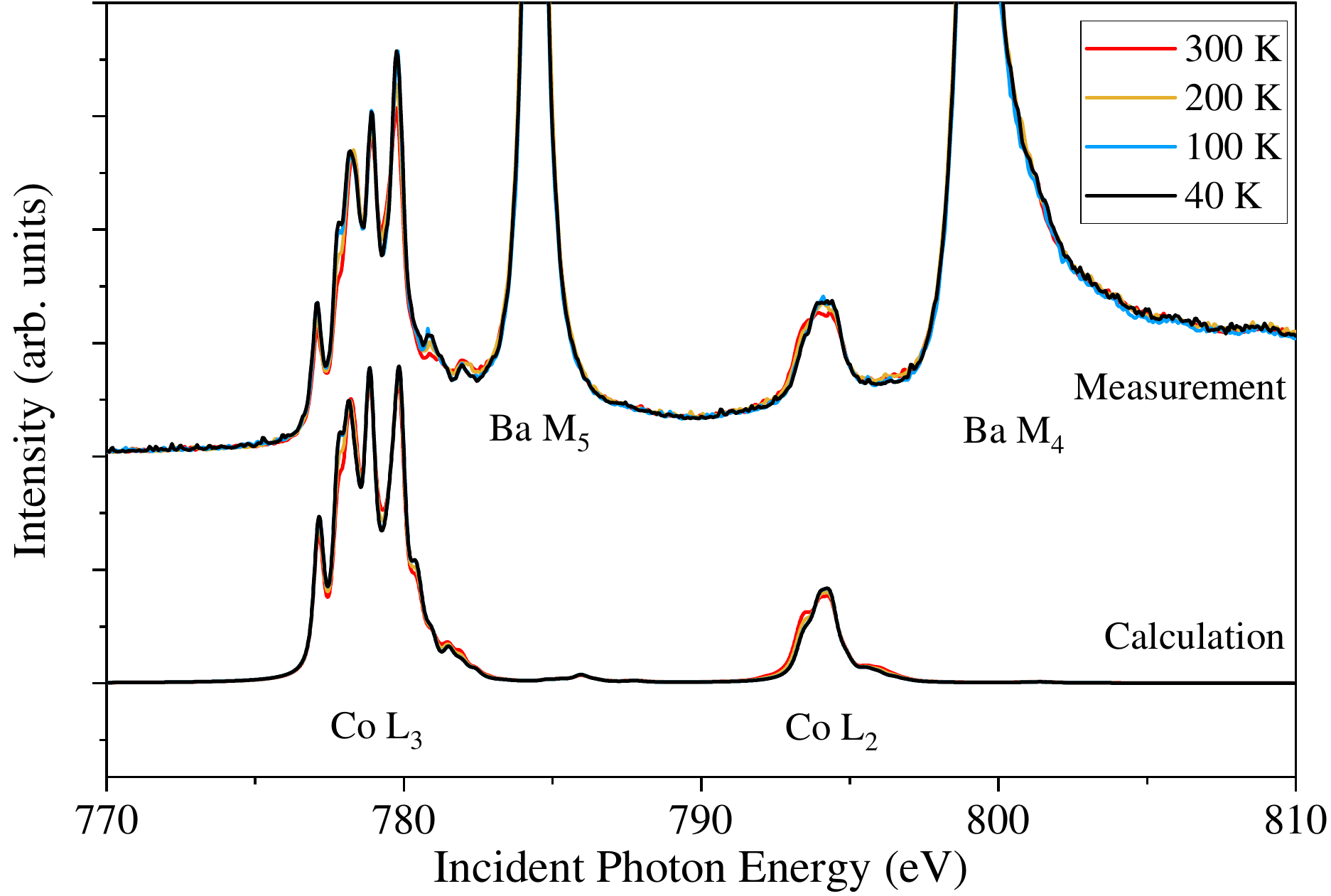}
    \caption{Experimental and calculated temperature dependence of isotropic spectra measured for Na$_2$BaCo(PO$_4$)$_2$}
    \label{fig:isospectra}
\end{figure}

The temperature dependence of the isotropic Co $L_{2,3}$ and Ba $M_{4,5}$ absorption spectra is presented in figure \ref{fig:isospectra}. These spectra were constructed using $(2 \times I_{\mathrm{LH}} + I_{\mathrm{LV}})/3$, where $I_{\mathrm{LH}}$ and $I_{\mathrm{LV}}$ denote inverse partial fluorescence yield (iPFY) spectra measured using linear horizontal (LH) and linear vertical (LV) polarization, respectively. The observed temperature dependence arises from  thermal population of low-lying excited states as temperature increases. This behavior is expected for a high-spin Co$^{2+}$ ion in octahedral coordination, where the lowest-energy states are split by 3$d$ SOC of ~66 meV. This relatively small energy scale allows for significant thermal population of low-lying excited states with a few hundred Kelvin, leading to the deviation of the Curie Weiss behavior in magnetic susceptibility \cite{burnusLocalElectronicStructure2008} and the strong temperature dependence of the Co $L_{2,3}$ XAS spectra.

The Co $L_{2,3}$ spectral shape is governed by a complex multiplet structure resulting from intra-atomic Coulomb interactions, local crystal fields, and hybridization with oxygen ligands. To extract more detailed information on the local electronic structure, we performed multiplet calculations based on a CoO$_6$ cluster with $D_{3d}$ symmetry with the $C_3$ axis aligned along the $z$-direction, consistent with the local Co point group symmetry of $\bar{3}m$. Input parameters for the calculations can be found in \cite{parameters}. We are able to accurately reproduce the spectra and its temperature dependence as shown in figure \ref{fig:isospectra}. Our simulations include the full atomic multiplet theory, incorporation of the 3$d$–3$d$ and 2$p$–3$d$ Coulomb interactions, SOC for both 2$p$ and 3$d$ electrons, Co 3$d$–O 2$p$ hybridization, and ligand crystal fields. 

\begin{figure}[h]
    \centering
    \includegraphics[width=\linewidth]{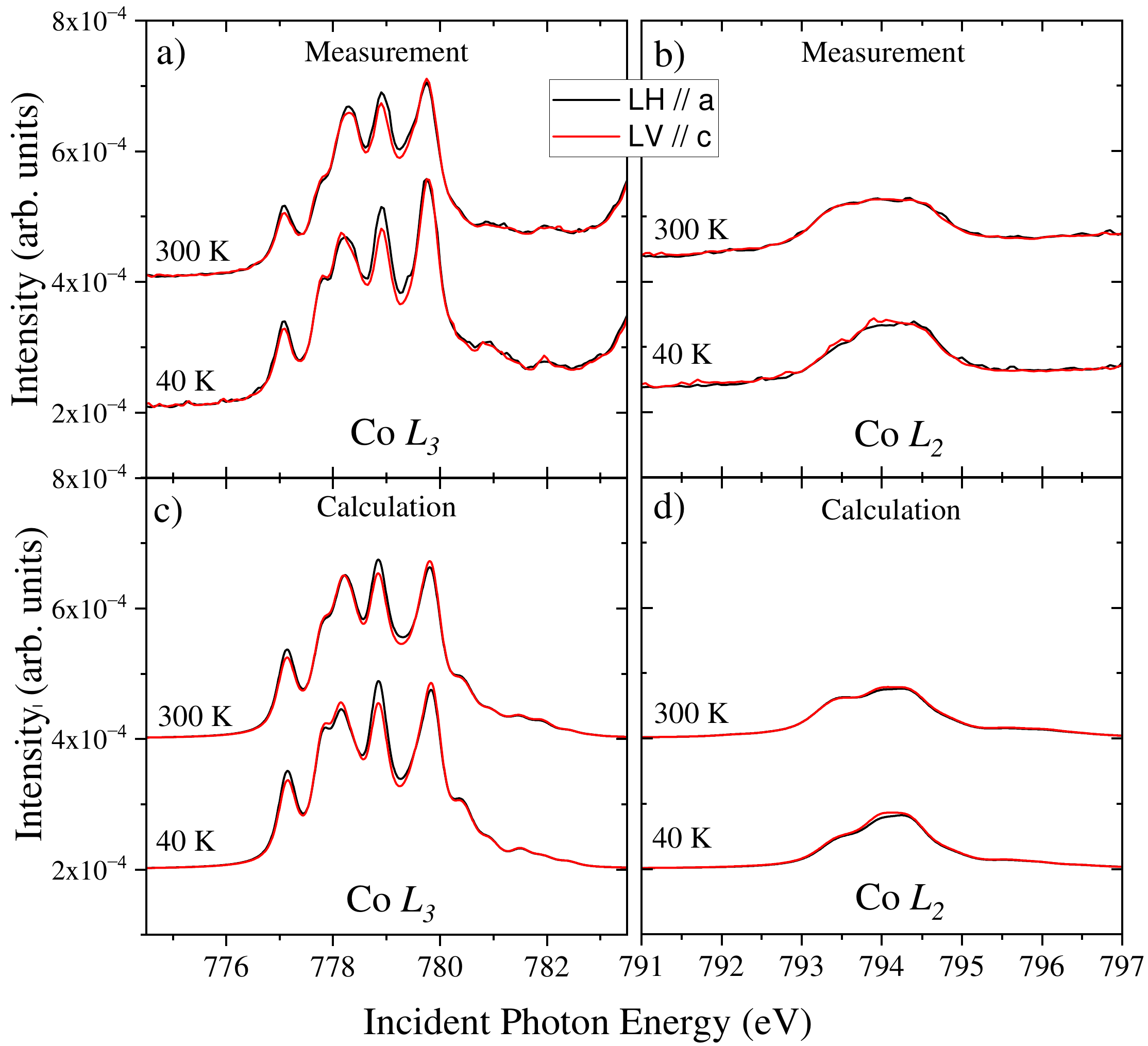}
    \caption{Polarization-dependent Co $L_{2,3}$ XAS spectra taken at 40 K and 300 K: (a,b) experiment and (c,d) calculation. Incident beam polarization is along either \textit{a} (LH) or along \textit{c} direction (LV).}
    \label{fig:LD_edges}
\end{figure}

\begin{figure}[h!]
    \centering
    \includegraphics[width=\linewidth]{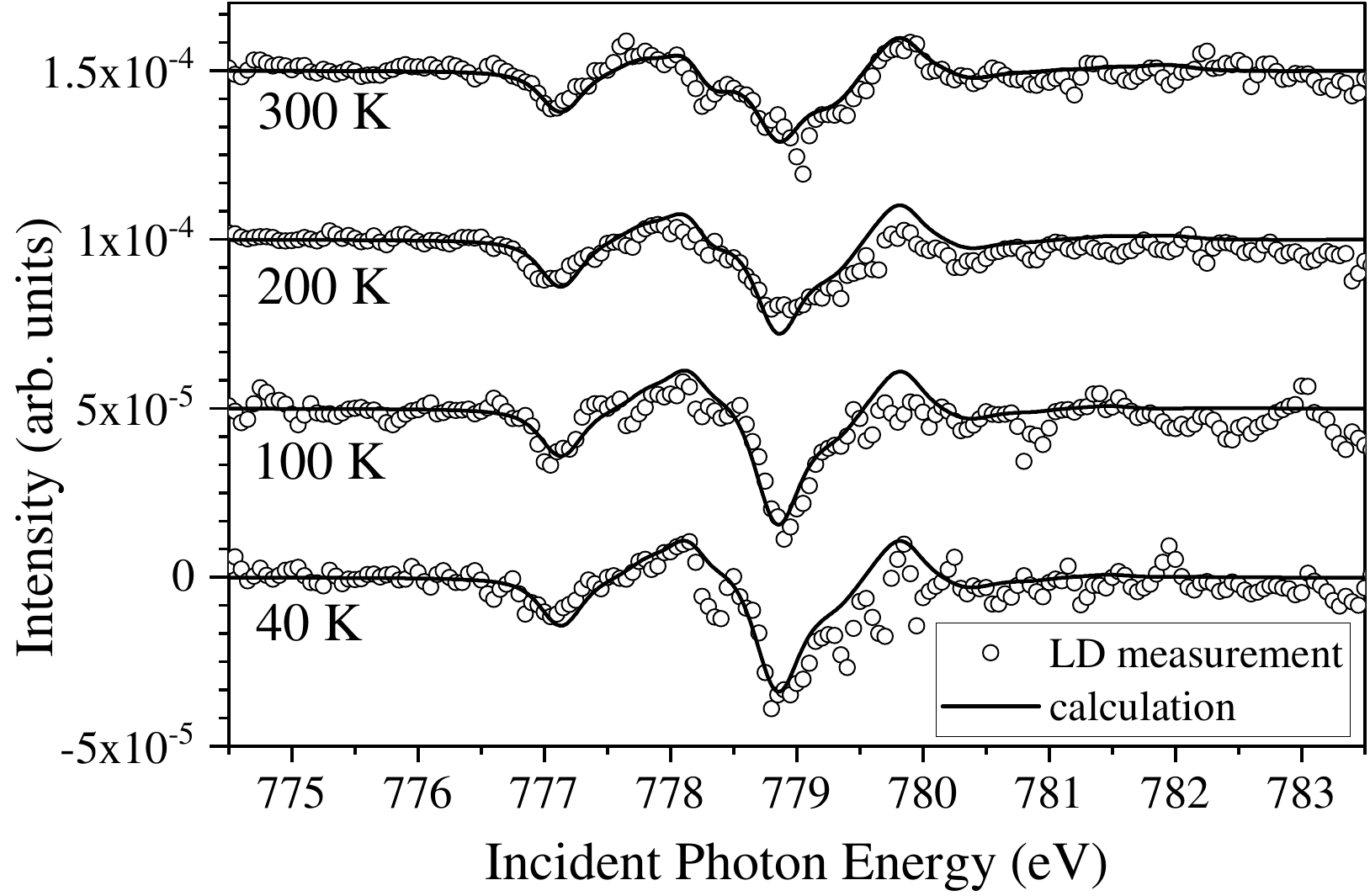}
    \caption{Temperature dependence of the measured and calculated linear dichroic spectra defined as $I_{LD} = I_{LV}-I_{LH} $ (see text for definition of the polarizations directions)}
    \label{fig:LD}
\end{figure}

Figure \ref{fig:LD_edges} a) and b) display the polarization-dependent Co $L_{2,3}$ XAS spectra taken at 40 K and 300 K.
The spectra taken with LV and LH polarizations are almost identical, indicating that the local environment is nearly cubic, close to ideal $O_h$ symmetry. This is further supported by our simulations, which match the measurements very well as shown in figure \ref{fig:LD_edges} c) and d). With the difference between the LV and LH polarizations being so small, we present in figure~\ref{fig:LD} with an enlarge vertical scale the linear dichroic spectra, defined as the difference between the spectra taken with LV and LH polarizations. We were also able to simulate these LD spectra accurately, including their temperature dependence. 

The best fit to the experimental data was achieved using a trigonal distortion parameter of \( D_{\mathrm{trig}} = E_{e_{g}^\pi} - E_{a_{1g}} = 8 \, \text{meV} \). This value is significantly smaller than the SOC constant energy scale (66\,meV), suggesting only minor deviations from a $ J_{\text{eff}} = \frac{1}{2} $ ground state (if only the $t_{2g}$ orbitals span the Hilbert space, see discussion section below). This agreement between calculation and measurement indicates that the present crystal field and hopping parameters accurately capture the local electronic structure of the Co$^{2+}$ ions in Na$_2$BaCo(PO$_4$)$_2$. \red{Here we note that the positive value of the trigonal distortion is a direct result of the observed polarization dependence of the XAS spectra, namely that the hole density in the $t_{2g}$ orbital manifold is more in the $e_g^{\pi}$ orbitals than in the $a_{1g}$. This finding may seem counter intuitive since the refined crystal structure shows a slight elongation along the $c$-axis of the local octahedra. However, longer-range interactions apparently contribute as well, thereby reversing the local crystal field scheme, as is the case in, for example, BaCoO$_3$ \cite{chinSpinorbitCouplingCrystalfield2019} and Na$_3$Co$_2$SbO$_6$ \cite{vanveenendaalElectronicStructureCo2023a}}. 

\section{Magnetic Susceptibility}

\begin{figure}
    \centering
    \includegraphics[width=\linewidth]{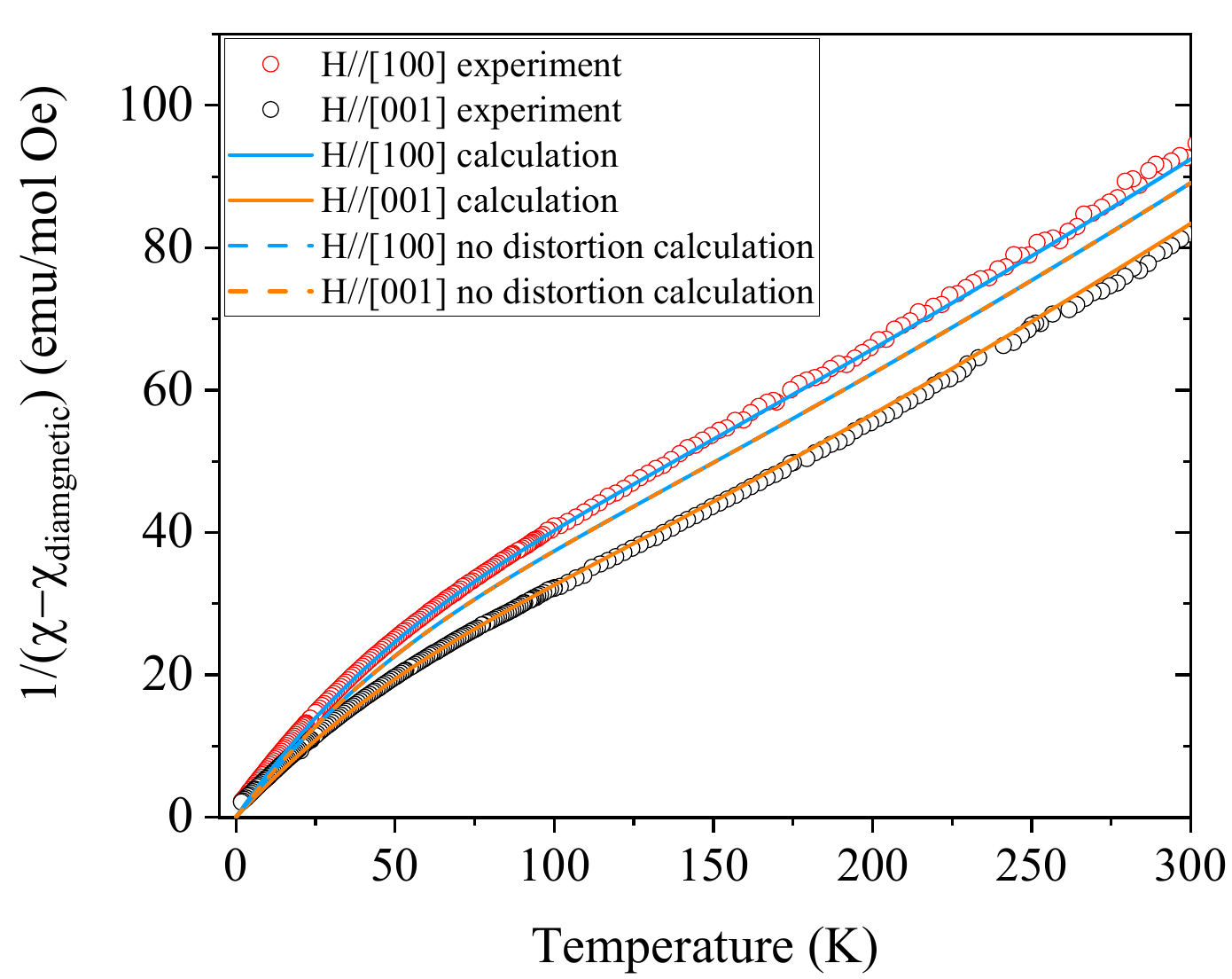}
    \caption{Inverse magnetic susceptibility, $\chi^{-1}$, measured under an applied field of 0.1 T set either along the [100] or [001] crystallographic directions, compared with calculations from our cluster model. Dashed lines indicate simulated susceptibilities obtained without including trigonal distortion in the calculations.}

    \label{fig:susc}
\end{figure}

To further check our model, we used the parameters obtained from our XAS analysis \cite{parameters} to simulate the magnetic susceptibility measurements with the applied field along different directions. Figure \ref{fig:susc} shows susceptibility data measured for a 0.1 T field applied along the [001] and [100] directions. The data are normalized to sample weight and applied magnetic field and corrected for diamagnetic contributions following \cite{bainDiamagneticCorrectionsPascals2008}. The experimental susceptibility clearly deviates from ideal Curie–Weiss behavior, as previously explained \cite{burnusLocalElectronicStructure2008}, due to the thermal population of low‐lying excited states. We are also able to simulate these curves with our cluster model as shown in figure \ref{fig:susc}. In order to calculate this temperature dependence, we start with the same parameters used for XAS, but now include the external magnetic field of 0.1 T along the relevant directions ([100] and [001] respectively).  We then calculate the expectation values of $\langle S\rangle$ and $\langle L\rangle$ for all the states. To calculate the magnetic susceptibility $\chi(T)$, at a given temperature $T$, we sum the contributions of each state $i$, weighted by their thermal occupation:

\begin{equation}
\chi \;=\;\frac{M}{H}
\;=\;
-\sum_i  \,
\frac{f_i(T)\langle L^{(i)} + 2S^{(i)}\rangle\,\mu_B}
{H_{\mathrm{Tesla}}}\frac{N_A}{10}
\label{eq:susc}
\end{equation}

With $f_i$ being the Boltzmann distribution for a given temperature \textit{T} and $Z$ the partition sum.
\[
f_i(T) \;=\; \frac{1}{Z} e^{-E_i / k_B T},
\qquad
Z \;=\; \sum_i e^{-E_i / k_B T}
\]
$H_{\mathrm{Tesla}}$ corresponds to the external applied magnetic field. To have the same units as measurement, $\mathrm{mol\,Oe/emu}$, we multiply by $\mu_B$, Bohr magneton, and $N_A$, the Avogadro number. \red{The $1/10$ factor comes from the conversion from Tesla to Gauss.} Figure \ref{fig:susc} also highlights the importance of the trigonal distortion parameter in the description of this system. Without this term, the calculated susceptibilities along the two crystallographic directions are identical, in contrast to the experimental results.

A very small antiferromagnetic exchange  (\red{$H_{\mathrm{ex}} = J \langle S \rangle$} $\approx$ -0.0018 meV) was applied to match the experiment. This weak exchange term can be interpreted as an emergent local field, generated by short‐range antiferromagnetic correlations under the influence of the applied magnetic field. A band structure study based on single‐crystal electron spin resonance, powder static magnetization and susceptibility measurements, reports a similarly small, isotropic antiferromagnetic exchange constant of \(J = 0.8\text{–}1.37\ \mathrm{K}\) (approximately \(0.068\text{–}0.118\   \  \mathrm{meV}\)) \cite{wellmFrustrationEnhancedKitaev2021}. This corroborates the picture of very weak, short‐range antiferromagnetic correlations superimposed on the dominant single‐ion crystal‐field and spin–orbit interactions. Despite its small magnitude, this local field is sufficient to fine‐tune the susceptibility curve, especially in the regime where thermal populations of excited states evolve rapidly.  Crucially, because $H_{\mathrm{ex}}$ couples only to the spin component it does not perturb the XAS lineshape. The resulting consistency between spectroscopic and bulk magnetic data therefore highlights the validity of our single‐cluster Hamiltonian parameters.

We can estimate effective g factors using the relation $M=gM_{J_{\text{eff}}}$, considering $M_{J_{\text{eff}}}=1/2$, to be $g_a$ = 4.54 and $g_c$ = 5.18 (values obtained in a calculation using H = 6 T). \red{Alternatively, using the relation $\Delta E = g \mu_B B$ where $\Delta E$ is the splitting of the doublet due to the magnetic field, we obtain $g_a$ = 4.30 and $g_c$ = 4.90}. The latter values are in very good agreement to previously reported values of $g_a$ =4.24 and  $g_c$ =4.83  ~\cite{wellmFrustrationEnhancedKitaev2021}. This small anisotropy can be related with the very small trigonal distortion present in the system. 

\section{Discussion}

\red{Since covalency play a significant role in transition metal oxides, local crystal fields are determined not only by the electrostatic potentials (“ionic” crystal fields) but very much also by the hybridization with the ligands. To determine these “effective” crystal fields, we need to calculate the differences in the total energies} of the states of interest with the SOC set to zero. To obtain the effective $10Dq_{\text{eff}}$, we first set $D_{trig}^{ion}=0$ and find the difference in energy between the $t_{2g}^5e^2_g$ configuration ground state $^4T_1$ term with the first excited state corresponding to a $t_{2g}^4e^3_g$ configuration $^4T_2$ term, giving 10$Dq_{\text{eff}}=$ 0.626 eV. To determine the effective trigonal distortion we do a similar calculation but this time including $D_{trig}^{ion}= 8$ meV, as found in our experimental fit. When a cubic state is distorted trigonally, the  previous $^4T_1$ ground state term will split into $^4A$  and $^4E$ states. The difference between this two states will give the effective trigonal distortion, which for the current set of parameters corresponds to 11 meV. \red{This value is in agreement with a recent Raman study \cite{mouComparativeRamanScattering2024} and supports the finding from a band structure study of Na$_2$BaCo(PO$_4$)$_2$ \cite{wellmFrustrationEnhancedKitaev2021}}. In our calculations, we obtain a separation between ground state $J_\mathrm{eff}=1/2$ and first excited state $J_\mathrm{eff}=3/2$ of about 40 meV. This value explains the excitations found by recent Raman scattering experiments: 39 meV \cite{mouComparativeRamanScattering2024} and 38 meV \cite{hussainExperimentalEvidenceCrystalfield2025a} .

In our calculations we include the hybridization of the Co $3d$ with neighbouring O $2p$ states, resulting in a ground state consisting of the following local configurations:

\begin{equation}
\Psi(\mathrm{Co}^{2+}) = \alpha_7 |d^7\rangle + \alpha_8 |d^8 \underline{L}\rangle + \alpha_9 |d^9\underline{L}^2\rangle
\end{equation}

where $\underline{L}$ denotes a ligand hole, and $\sum_{i=7}^9 \alpha_i^2=1$. We obtain the following weights for each configuration: $\alpha_7^2 = 0.886$, $\alpha_8^2 = 0.112$ and  $\alpha_9^2 = 0.0023 $. We thus have in total 2.88 holes in the $3d$ shell with a dominant $d^7$ configuration for the ground state.

The dominant $d^7$ ground state combined with the very small trigonal distortion further hints that this compound should be very close to ideal $J_{\text{eff}}=1/2$ ground state. To quantify this, we computed the expectation value $\langle \mathbf{J}_{\mathrm{eff}}^2 \rangle$ by constructing the effective total angular momentum operator $\mathbf{J}_{\mathrm{eff}} = \mathbf{L}_{\mathrm{eff}} + \mathbf{S}$, where the effective orbital angular momentum operator $\mathbf{L}_{\mathrm{eff}}$ is obtained by rotating the standard angular momentum operator $\mathbf{L}$ into the cubic harmonics basis \cite{agrestiniProbingEff02018}. The rotation matrix is modified to retain only the $ t_{2g}$ subset of the 3\textit{d} orbital manifold. After projecting out the  $e_g$  orbitals, the angular momentum operator is then rotated back into the spherical harmonics basis. This total angular momentum operator is then evaluated over the full CoO$_6$ cluster to include the effects of Co-O covalency.

It is important to note that a value of $J_{\text{eff}}=1/2$ is only obtained when $t_{2g}$ orbitals fully span the Hilbert space, one can do this by setting 10\textit{Dq} to a infinitely large value. For example, setting 10$Dq$ = 10 eV one finds $J_{\mathrm{eff}} =$ 0.501. For the found 10$Dq$ crystal field, not in the infinite limit, we obtain $J_{\mathrm{eff}} =$ 0.854, indicating some participation of $e_g$ orbitals. Nevertheless, the ground‐state remains a well‐separated Kramers doublet, preserving its quasi–$J_{\mathrm{eff}} = \tfrac12$ character.

\section{Conclusion}

In this work, we have demonstrated that temperature‐dependent X-ray absorption spectroscopy offers an exceptionally powerful and precise probe of the local electronic structure in highly insulating transition‐metal compounds. By directly measuring the dichroic spectra of Na$_2$BaCo(PO$_4$)$_2$ as a function of temperature, we observe a very small distortion of the CoO$_6$ of $D_\mathrm{trig}\approx11$ meV. We also confirm a Co\(^{2+}\) valence state, with the predominance of a \(J_{\mathrm{eff}} = \tfrac12\) ground state in Na$_2$BaCo(PO$_4$)$_2$, with our model also being able to reproduce rather well magnetic susceptibility measurements along different directions, highlighting the local geometric frustration experienced by the Co\(^{2+}\) ions.

The crystal‐field parameters and other relevant electronic structure parameters determined here should provide a solid quantitative foundation for future theoretical descriptions of Na$_2$BaCo(PO$_4$)$_2$ and related two‐dimensional \(J_{\mathrm{eff}}=\tfrac12\) triangular‐lattice magnets.  More broadly, the ability of temperature‐dependent XAS to resolve meV scale distortions and orbital occupations offers a powerful toolkit for understanding the microscopic interactions that give rise to unconventional magnetic phenomena in highly frustrated systems.   


\begin{acknowledgments}

M.M.F.-C. greatly acknowledges funding from the  Deutsche Forschungsgemeinschaft (DFG, German Research Foundation) Grant No. 387555779. Work at MPI-CPfS Dresden was partially supported by SFB1143 (Project No. 247310070). The authors acknowledge the support from the Max Planck-POSTECH-Hsinchu Center for Complex Phase Materials.

\end{acknowledgments}
\section*{Data Availability}

The data that support the findings of this article are openly
available \cite{carvalhoEdmondXe7bis2025}.

\balance
\bibliographystyle{unsrt}
\bibliography{references}

@article{ANDERSON1973153,
  title = {{Resonating Valence Bonds: {{A}} New Kind of Insulator?}},
  author = {Anderson, P.W.},
  year = {1973},
  journal= {Materials Research Bulletin},
  volume = {8},
  number = {2},
  pages = {153--160},
  issn = {0025-5408},
  doi = {10.1016/0025-5408(73)90167-0},
}

@article{bernuNeelOrderSpin1993,
  title = {{N\'eel Order versus Spin Liquid in Quantum {Heisenberg} Antiferromagnets on Triangular and Kagom\'e Lattices}},

  author = {Bernu, B. and Lecheminant, P. and Lhuillier, C. and Pierre, L.},
  year = {1993},
  month = jan,
  journal = {Physica Scripta},
  volume = {1993},
  number = {T49A},
  pages = {192},
  issn = {1402-4896},
  doi = {10.1088/0031-8949/1993/T49A/032},

}

@article{bernuSignatureNeelOrder1992,
  title = {{Signature of {N{\'e}el} Order in Exact Spectra of Quantum Antiferromagnets on Finite Lattices}}
,
  author = {Bernu, B. and Lhuillier, C. and Pierre, L.},
  year = {1992},
  month = oct,
  journal = {Physical Review Letters},
  volume = {69},
  number = {17},
  pages = {2590--2593},
  publisher = {American Physical Society},
  doi = {10.1103/PhysRevLett.69.2590},
  
}

@article{bernuExactSpectraSpin1994,
  title = {{Exact Spectra, Spin Susceptibilities, and Order Parameter of the Quantum {{Heisenberg}} Antiferromagnet on the Triangular Lattice}},
  author = {Bernu, B. and Lecheminant, P. and Lhuillier, C. and Pierre, L.},
  year = {1994},
  month = oct,
  journal = {Physical Review B},
  volume = {50},
  number = {14},
  pages = {10048--10062},
  publisher = {American Physical Society},
  doi = {10.1103/PhysRevB.50.10048},
  
}

@article{whiteNeelOrderSquare2007,
  title = {{Ne{\`e}l {Order} in {Square} and {Triangular Lattice Heisenberg Models}}},

  author = {White, Steven R. and Chernyshev, A. L.},
  year = {2007},
  month = sep,
  journal = {Physical Review Letters},
  volume = {99},
  number = {12},
  pages = {127004},
  publisher = {American Physical Society},
  doi = {10.1103/PhysRevLett.99.127004},
  
}

@article{bauerSchwingerbosonMeanfieldStudy2017,
  title = {{Schwinger-Boson Mean-Field Study of the $J_1\text{-}J_2$ {Heisenberg} Quantum Antiferromagnet on the Triangular Lattice}}
,
  author = {Bauer, Dag-Vidar and Fj{\ae}restad, J. O.},
  year = {2017},
  month = oct,
  journal = {Physical Review B},
  volume = {96},
  number = {16},
  pages = {165141},
  publisher = {American Physical Society},
  doi = {10.1103/PhysRevB.96.165141},
  
}

@article{ferrariDynamicalStructureFactor2019,
  title = {{Dynamical {Structure Factor} of the $J_1\text{-}J_2$ {Heisenberg Model} on the {Triangular Lattice}: {Magnons}, {Spinons}, and {Gauge Fields}}},


  author = {Ferrari, Francesco and Becca, Federico},
  year = {2019},
  month = aug,
  journal = {Physical Review X},
  volume = {9},
  number = {3},
  pages = {031026},
  publisher = {American Physical Society},
  doi = {10.1103/PhysRevX.9.031026},
  
}

@article{gongGlobalPhaseDiagram2017,
 title = {{Global Phase Diagram and Quantum Spin Liquids in a Spin-$\frac{1}{2}$ Triangular Antiferromagnet}},
  author = {Gong, Shou-Shu and Zhu, W. and Zhu, J.-X. and Sheng, D. N. and Yang, Kun},
  year = {2017},
  month = aug,
  journal = {Physical Review B},
  volume = {96},
  number = {7},
  pages = {075116},
  publisher = {American Physical Society},
  doi = {10.1103/PhysRevB.96.075116},
  
}

@article{huDiracSpinLiquid2019,
  title = {{Dirac {Spin Liquid} on the Spin-$\frac{1}{2}$ {Triangular Heisenberg Antiferromagnet}}},
  author = {Hu, Shijie and Zhu, W. and Eggert, Sebastian and He, Yin-Chen},
  year = {2019},
  month = nov,
  journal = {Physical Review Letters},
  volume = {123},
  number = {20},
  pages = {207203},
  publisher = {American Physical Society},
  doi = {10.1103/PhysRevLett.123.207203}
 
}

@article{iqbalSpinLiquidNature2016,
  title = {{Spin Liquid Nature in the {Heisenberg} $J_1\text{-}J_2$ Triangular Antiferromagnet}},

  author = {Iqbal, Yasir and Hu, Wen-Jun and Thomale, Ronny and Poilblanc, Didier and Becca, Federico},
  year = {2016},
  month = apr,
  journal = {Physical Review B},
  volume = {93},
  number = {14},
  pages = {144411},
  publisher = {American Physical Society},
  doi = {10.1103/PhysRevB.93.144411},
 

}

@article{maksimovAnisotropicExchangeMagnetsTriangular2019,
  title = {{Anisotropic-Exchange Magnets on a Triangular Lattice: Spin Waves, Accidental Degeneracies, and Dual Spin Liquids}},
  author = {Maksimov, P. A. and Zhu, Zhenyue and White, Steven R. and Chernyshev, A. L.},
  year = {2019},
  month = apr,
  journal = {Physical Review X},
  volume = {9},
  number = {2},
  pages = {021017},
  publisher = {American Physical Society},
  doi = {10.1103/PhysRevX.9.021017}
  
  
}

@article{saadatmandDetectionCharacterizationSymmetrybroken2017,
  title = {{Detection and Characterization of Symmetry-Broken Long-Range Orders in the Spin-$\frac{1}{2}$ Triangular {Heisenberg} Model}},


  author = {Saadatmand, S. N. and McCulloch, I. P.},
  year = {2017},
  month = aug,
  journal = {Physical Review B},
  volume = {96},
  number = {7},
  pages = {075117},
  publisher = {American Physical Society},
  doi = {10.1103/PhysRevB.96.075117},
  urldate = {2025-07-01},
 
}

@article{zhuSpinLiquidPhase2015,
  title = {{Spin Liquid Phase of the $S=\frac{1}{2} \phantom{\rule{4.pt}{0ex}} J_1\!-\!J_2$ {Heisenberg} Model on the Triangular Lattice}},

  author = {Zhu, Zhenyue and White, Steven R.},
  year = {2015},
  month = jul,
  journal = {Physical Review B},
  volume = {92},
  number = {4},
  pages = {041105},
  publisher = {American Physical Society},
  doi = {10.1103/PhysRevB.92.041105},
  urldate = {2025-07-01},

}

@article{zhuTopographySpinLiquids2018,
  title = {{Topography of {{Spin Liquids}} on a {{Triangular Lattice}}}},
  author = {Zhu, Zhenyue and Maksimov, P. A. and White, Steven R. and Chernyshev, A. L.},
  year = {2018},
  month = may,
  journal = {Physical Review Letters},
  volume = {120},
  number = {20},
  pages = {207203},
  publisher = {American Physical Society},
  doi = {10.1103/PhysRevLett.120.207203},
  urldate = {2025-07-01},
 
}

@article{liSpinLiquidsGeometrically2020,
  title = {{Spin Liquids in Geometrically Perfect Triangular Antiferromagnets}},
  author = {Li, Yuesheng and Gegenwart, Philipp and Tsirlin, Alexander A},
  year = {2020},
  month = mar,
  journal = {Journal of Physics: Condensed Matter},
  volume = {32},
  number = {22},
  pages = {224004},
  publisher = {IOP Publishing},
  issn = {0953-8984},
  doi = {10.1088/1361-648X/ab724e},
  urldate = {2025-07-01},
  langid = {english},
 
}

@article{liPossibleItinerantExcitations2020,
  title = {{Possible Itinerant Excitations and Quantum Spin State Transitions in the Effective Spin-1/2 Triangular-Lattice Antiferromagnet {{Na$_2$BaCo}}({{PO$_4$}})$_2$}},
  author = {Li, N. and Huang, Q. and Yue, X. Y. and Chu, W. J. and Chen, Q. and Choi, E. S. and Zhao, X. and Zhou, H. D. and Sun, X. F.},
  year = {2020},
  month = aug,
  journal = {Nature Communications},
  volume = {11},
  number = {1},
  pages = {4216},
  publisher = {Nature Publishing Group},
  issn = {2041-1723},
  doi = {10.1038/s41467-020-18041-3},
  urldate = {2025-07-01},
  copyright = {2020 The Author(s)},
  langid = {english},
  
}

@article{zhongStrongQuantumFluctuations2019,
  title = {{Strong Quantum Fluctuations in a Quantum Spin Liquid Candidate with a {{Co-based}} Triangular Lattice}},
  author = {Zhong, Ruidan and Guo, Shu and Xu, Guangyong and Xu, Zhijun and Cava, Robert J.},
  year = {2019},
  month = jul,
  journal = {Proceedings of the National Academy of Sciences},
  volume = {116},
  number = {29},
  pages = {14505--14510},
  publisher = {Proceedings of the National Academy of Sciences},
  doi = {10.1073/pnas.1906483116},
  urldate = {2025-07-01}
}

@article{leeTemporalFieldEvolution2021,
  title = {{Temporal and Field Evolution of Spin Excitations in the Disorder-Free Triangular Antiferromagnet $\text{Na}_2\text{BaCo}(\text{PO}_4)_2$}},

  author = {Lee, S. and Lee, C. H. and Berlie, A. and Hillier, A. D. and Adroja, Devashibhai T. and Zhong, Ruidan and Cava, R. J. and Jang, Z. H. and Choi, K.-Y.},
  year = {2021},
  month = jan,
  journal = {Physical Review B},
  volume = {103},
  number = {2},
  pages = {024413},
  publisher = {American Physical Society},
  doi = {10.1103/PhysRevB.103.024413},
  urldate = {2025-07-01}
}

@article{huangThermalConductivityTriangularlattice2022,
  title = {{Thermal Conductivity of Triangular-Lattice Antiferromagnet {{Na$_2$BaCo}}({{PO$_4$}})$_2$: {{Absence}} of Itinerant Fermionic Excitations}},

  author = {Huang, Y. Y. and Dai, D. Z. and Zhao, C. C. and Ni, J. M. and Wang, L. S. and Pan, B. L. and Gao, B. and Dai, Pengcheng and Li, S. Y.},
  year = {2022},
  month = jun,
  eprint = {2206.08866},
  primaryclass = {cond-mat},
  publisher = {arXiv},
  journal = {10.48550/arXiv.2206.08866},
  doi = {10.48550/arXiv.2206.08866},
  urldate = {2025-07-01},
  archiveprefix = {arXiv}
}

@article{gaoSpinSupersolidityNearly2022,
  title = {{Spin Supersolidity in Nearly Ideal Easy-Axis Triangular Quantum Antiferromagnet {{Na$_2$BaCo}}({{PO$_4$}})$_2$}},
  author = {Gao, Yuan and Fan, Yu-Chen and Li, Han and Yang, Fan and Zeng, Xu-Tao and Sheng, Xian-Lei and Zhong, Ruidan and Qi, Yang and Wan, Yuan and Li, Wei},
  year = {2022},
  month = sep,
  journal = {npj Quantum Materials},
  volume = {7},
  number = {1},
  pages = {89},
  publisher = {Nature Publishing Group},
  issn = {2397-4648},
  doi = {10.1038/s41535-022-00500-3},
  urldate = {2025-07-01},
  copyright = {2022 The Author(s)},
  langid = {english}
}

@article{xiangGiantMagnetocaloricEffect2024,
  title = {{Giant Magnetocaloric Effect in Spin Supersolid Candidate {{Na$_2$BaCo}}({{PO$_4$}})$_2$}},
  author = {Xiang, Junsen and Zhang, Chuandi and Gao, Yuan and Schmidt, Wolfgang and Schmalzl, Karin and Wang, Chin-Wei and Li, Bo and Xi, Ning and Liu, Xin-Yang and Jin, Hai and Li, Gang and Shen, Jun and Chen, Ziyu and Qi, Yang and Wan, Yuan and Jin, Wentao and Li, Wei and Sun, Peijie and Su, Gang},
  date = {2024-01},
year =2024,
month = jan ,
 journal = {Nature},
  volume = {625},
  number = {7994},
  pages = {270--275},
  publisher = {Nature Publishing Group},
  issn = {1476-4687},
  doi = {10.1038/s41586-023-06885-w},
  url = {https://www.nature.com/articles/s41586-023-06885-w},
  urldate = {2025-07-01},
  langid = {english}
}

@article{chubukovQuantumTheoryAntiferromagnet1991,
  title = {{Quantum Theory of an Antiferromagnet on a Triangular Lattice in a Magnetic Field}},
  author = {Chubukov, A. V. and Golosov, D. I.},
  year = {1991},
  month = jan,
  journal = {Journal of Physics: Condensed Matter},
  volume = {3},
  number = {1},
  pages = {69},
  issn = {0953-8984},
  doi = {10.1088/0953-8984/3/1/005},
  urldate = {2025-07-01},
  langid = {english}
}

@article{starykhUnusualOrderedPhases2015,
  title = {{Unusual Ordered Phases of Highly Frustrated Magnets: A Review}},

  author = {Starykh, Oleg A},
  year = {2015},
  month = apr,
  journal = {Reports on Progress in Physics},
  volume = {78},
  number = {5},
  pages = {052502},
  publisher = {IOP Publishing},
  issn = {0034-4885},
  doi = {10.1088/0034-4885/78/5/052502},
  urldate = {2025-07-01},
  langid = {english}
}

@article{woodlandContinuumExcitationsSharp2025,
  title = {From Continuum Excitations to Sharp Magnons via Transverse Magnetic Field in the Spin-$\frac{1}{2}$ {{Ising-like}} Triangular Lattice Antiferromagnet ${\text{{{Na}}}}_{ 2}\text{{{BaCo}}}{ ({\text{{{PO}}}}_{4})}_{2}$},
  author = {Woodland, Leonie and Okuma, Ryutaro and Stewart, J. Ross and Balz, Christian and Coldea, Radu},
  year = 2025,
  month = sep,
  journal = {Physical Review B},
  volume = {112},
  number = {10},
  pages = {104413},
  publisher = {American Physical Society},
  doi = {10.1103/1pvl-kzjm}
}

@article{shengContinuumSpinExcitations2025,
  title = {{Continuum of Spin Excitations in an Ordered Magnet}},
  author = {Sheng, Jieming and Wang, Le and Jiang, Wenrui and Ge, Han and Zhao, Nan and Li, Tiantian and Kofu, Maiko and Yu, Dehong and Zhu, Wei and Mei, Jia-Wei and Wang, Zhentao and Wu, Liusuo},
  year = {2025},
month = apr,
  journal = {The Innovation},
  volume = {6},
  number = {4},
  publisher = {The Innovation},
  issn = {2666-6758},
  urldate = {2025-07-01},
  copyright = {http://creativecommons.org/licenses/by/3.0/},
  langid = {english}
}

@article{burnusLocalElectronicStructure2008,
  title = {{Local Electronic Structure and Magnetic Properties of {{LaMn}}$_{0.5}${{Co}}$_{0.5}${{O}}$_3$ Studied by x-Ray Absorption and Magnetic Circular Dichroism Spectroscopy}},
  author = {Burnus, T. and Hu, Z. and Hsieh, H. H. and Joly, V. L. J. and Joy, P. A. and Haverkort, M. W. and Wu, Hua and Tanaka, A. and Lin, H.-J. and Chen, C. T. and Tjeng, L. H.},
  year = {2008},
  month = mar,
  journal = {Physical Review B},
  volume = {77},
  number = {12},
  pages = {125124},
  issn = {1098-0121, 1550-235X},
  doi = {10.1103/PhysRevB.77.125124},
  urldate = {2025-07-02},
  copyright = {http://link.aps.org/licenses/aps-default-license},
  langid = {english}
}

@article{achkarBulkSensitiveXray2011,
  title = {{Bulk Sensitive X-Ray Absorption Spectroscopy Free of Self-Absorption Effects}},
  author = {Achkar, A. J. and Regier, T. Z. and Wadati, H. and Kim, Y.-J. and Zhang, H. and Hawthorn, D. G.},
  year = {2011},
  month = feb,
  journal = {Physical Review B},
  volume = {83},
  number = {8},
  pages = {081106},
  issn = {1098-0121, 1550-235X},
  doi = {10.1103/PhysRevB.83.081106},
  urldate = {2025-07-02},
  copyright = {http://link.aps.org/licenses/aps-default-license},
  langid = {english}
}

@article{achkarDeterminationTotalXray2011,
  title = {{Determination of Total X-Ray Absorption Coefficient Using Non-Resonant x-Ray Emission}},
  author = {Achkar, A. J. and Regier, T. Z. and Monkman, E. J. and Shen, K. M. and Hawthorn, D. G.},
  year = {2011},
  month = dec,
  journal = {Scientific Reports},
  volume = {1},
  number = {1},
  pages = {182},
  issn = {2045-2322},
  doi = {10.1038/srep00182},
  urldate = {2025-07-02},
  langid = {english}
}

@article{burnusValenceSpinOrbital2006,
  title = {{Valence, Spin, and Orbital State of {{Co}} Ions in One-Dimensional {{Ca}}$_3${{Co}}$_2${{O}}$_6$ : {{An}} x-Ray Absorption and Magnetic Circular Dichroism Study}},
  author = {Burnus, T. and Hu, Z. and Haverkort, M. W. and Cezar, J. C. and Flahaut, D. and Hardy, V. and Maignan, A. and Brookes, N. B. and Tanaka, A. and Hsieh, H. H. and Lin, H.-J. and Chen, C. T. and Tjeng, L. H.},
  year = {2006},
  month = dec,
  journal = {Physical Review B},
  volume = {74},
  number = {24},
  pages = {245111},
  issn = {1098-0121, 1550-235X},
  doi = {10.1103/PhysRevB.74.245111},
  urldate = {2025-07-02},
  copyright = {http://link.aps.org/licenses/aps-default-license},
  langid = {english}
}

@article{chinSpinorbitCouplingCrystalfield2019,
  title = {{Spin-Orbit Coupling and Crystal-Field Distortions for a Low-Spin 3d$^5$ State in {{BaCoO}}$_3$}},
  author = {Chin, Y. Y. and Hu, Z. and Lin, H.-J. and Agrestini, S. and Weinen, J. and Martin, C. and H{\'e}bert, S. and Maignan, A. and Tanaka, A. and Cezar, J. C. and Brookes, N. B. and Liao, Y.-F. and Tsuei, K.-D. and Chen, C. T. and Khomskii, D. I. and Tjeng, L. H.},
  year = {2019},
  month = nov,
  journal = {Physical Review B},
  volume = {100},
  number = {20},
  pages = {205139},
  issn = {2469-9950, 2469-9969},
  doi = {10.1103/PhysRevB.100.205139},
  urldate = {2025-07-02},
  langid = {english}
}

@article{degrootXrayAbsorptionDichroism1994,
  title = {{X-Ray Absorption and Dichroism of Transition Metals and Their Compounds}},
  author = {De Groot, F.M.F.},
  year = {1994},
  month = aug,
  journal = {Journal of Electron Spectroscopy and Related Phenomena},
  volume = {67},
  number = {4},
  pages = {529--622},
  issn = {03682048},
  doi = {10.1016/0368-2048(93)02041-J},
  urldate = {2025-07-02},
  copyright = {https://www.elsevier.com/tdm/userlicense/1.0/},
  langid = {english}
}

@article{linLocalOrbitalOccupation2010,
  title = {{Local Orbital Occupation and Energy Levels of {{Co}} in {{Na}}$_\text{x}${{CoO}}$_2$ : {{A}} Soft x-Ray Absorption Study}},
  author = {Lin, H.-J. and Chin, Y. Y. and Hu, Z. and Shu, G. J. and Chou, F. C. and Ohta, H. and Yoshimura, K. and H{\'e}bert, S. and Maignan, A. and Tanaka, A. and Tjeng, L. H. and Chen, C. T.},
  year = {2010},
  month = mar,
  journal = {Physical Review B},
  volume = {81},
  number = {11},
  pages = {115138},
  issn = {1098-0121, 1550-235X},
  doi = {10.1103/PhysRevB.81.115138},
  urldate = {2025-07-02},
  copyright = {http://creativecommons.org/licenses/by/3.0/},
  langid = {english}
}

@article{tanakaResonant3d3p1994,
  title = {{Resonant 3d, 3p and 3s {{Photoemission}} in {{Transition Metal Oxides Predicted}} at 2p {{Threshold}}}},
  author = {Tanaka, Arata and Jo, Takeo},
  year = {1994},
  month = jul,
  journal = {Journal of the Physical Society of Japan},
  volume = {63},
  number = {7},
  pages = {2788--2807},
  publisher = {The Physical Society of Japan},
  issn = {0031-9015},
  doi = {10.1143/JPSJ.63.2788},
  urldate = {2025-07-02}
}

@article{tsaiSubmicronSoftXray2019a,
  title = {{A Submicron Soft X-Ray Active Grating Monochromator Beamline for Ultra-High Resolution Angle-Resolved Photoemission Spectroscopy}},
  author = {Tsai, Huang-Ming and Fu, Huang-Wen and Kuo, Chang-Yang and Huang, Liang-Jen and Lee, Chang-Sheng and Hua, Chih-Yu and Kao, Kai-Yang and Lin, Hong-Ji and Fung, Hok-Sum and Chung, Shih-Chun and Chang, Chun-Fu and Chainani, Ashish and Tjeng, Liu Hao and Chen, Chien-Te},
  year = {2019},
  month = jan,
  journal = {AIP Conference Proceedings},
  volume = {2054},
  number = {1},
  pages = {060047},
  issn = {0094-243X},
  doi = {10.1063/1.5084678},
  urldate = {2025-07-02}
}

@article{haverkortBandsResonancesEdge2014,
  title = {{Bands, Resonances, Edge Singularities and Excitons in Core Level Spectroscopy Investigated within the Dynamical Mean-Field Theory}},
  author = {Haverkort, M. W. and Sangiovanni, G. and Hansmann, P. and Toschi, A. and Lu, Y. and Macke, S.},
  year = {2014},
  month = dec,
  journal = {Europhysics Letters},
  volume = {108},
  number = {5},
  pages = {57004},
  publisher = {{EDP Sciences, IOP Publishing and Societ{\`a} Italiana di Fisica}},
  issn = {0295-5075},
  doi = {10.1209/0295-5075/108/57004},
  urldate = {2025-07-02},
  langid = {english}
}

@article{haverkortMultipletLigandfieldTheory2012,
  title = {{Multiplet Ligand-Field Theory Using {{Wannier}} Orbitals}},
  author = {Haverkort, M. W. and Zwierzycki, M. and Andersen, O. K.},
  year = {2012},
  month = apr,
  journal = {Physical Review B},
  volume = {85},
  number = {16},
  pages = {165113},
  issn = {1098-0121, 1550-235X},
  doi = {10.1103/PhysRevB.85.165113},
  urldate = {2025-07-02},
  copyright = {http://link.aps.org/licenses/aps-default-license},
  langid = {english}
}

@article{luEfficientRealfrequencySolver2014,
  title = {{Efficient Real-Frequency Solver for Dynamical Mean-Field Theory}},
  author = {Lu, Y. and H{\"o}ppner, M. and Gunnarsson, O. and Haverkort, M. W.},
  year = {2014},
  month = aug,
  journal = {Physical Review B},
  volume = {90},
  number = {8},
  pages = {085102},
  issn = {1098-0121, 1550-235X},
  doi = {10.1103/PhysRevB.90.085102},
  urldate = {2025-07-02},
  copyright = {http://link.aps.org/licenses/aps-default-license},
  langid = {english}
}

@article{koepernikFullpotentialNonorthogonalLocalorbital1999,
  title = {{Full-Potential Nonorthogonal Local-Orbital Minimum-Basis Band-Structure Scheme}},
  author = {Koepernik, Klaus and Eschrig, Helmut},
  year = {1999},
  month = jan,
  journal = {Physical Review B},
  volume = {59},
  number = {3},
  pages = {1743--1757},
  issn = {0163-1829, 1095-3795},
  doi = {10.1103/PhysRevB.59.1743},
  urldate = {2025-07-02},
  copyright = {http://link.aps.org/licenses/aps-default-license},
  langid = {english}
}

@misc{fplo21,
  title        = {{FPLO} version 21.00},
  howpublished = {\url{https://www.fplo.de}},

}

@misc{Quanty,
    title = {{Quanty version 0.81}},
    howpublished = {\url{https://www.quanty.org}},

}

@article{wellmFrustrationEnhancedKitaev2021,
  title  = {{Frustration Enhanced by {Kitaev} Exchange in a $\tilde{j}_{\text{eff}}=\frac{1}{2}$ Triangular Antiferromagnet}},
  author = {Wellm, C. and Roscher, W. and Zeisner, J. and Alfonsov, A. and Zhong, R. and Cava, R. J. and Savoyant, A. and Hayn, R. and {van den Brink}, J. and B{\"u}chner, B. and Janson, O. and Kataev, V.},
  year = {2021},
  month = sep,
  journal = {Physical Review B},
  volume = {104},
  number = {10},
  pages = {L100420},
  publisher = {American Physical Society},
  doi = {10.1103/PhysRevB.104.L100420},
  urldate = {2025-07-03}
}

@article{bainDiamagneticCorrectionsPascals2008,
  title = {{Diamagnetic {{Corrections}} and {{Pascal}}'s {{Constants}}}},
  author = {Bain, Gordon A. and Berry, John F.},
  year = {2008},
  month = apr,
  journal = {Journal of Chemical Education},
  volume = {85},
  number = {4},
  pages = {532},
  issn = {0021-9584, 1938-1328},
  doi = {10.1021/ed085p532},
  urldate = {2025-07-03},
  langid = {english}
}

@misc{parameters,
  note = {\textit{U$_{dd}$} = 6.5 eV, \textit{U$_{pd}$} = 8.2 eV, charge energy transfer $\Delta$ = 6.5 eV, SOC = 0.066 eV, ionic crystal field 10$Dq^{ion}$ = 0.395 eV, $D^{ion}_{trig}$ = 8 meV, hybridization $V(e_{g}^{\sigma})$ = 2.45 eV, $V(e_{g}^{\pi})$ = 1.31 eV, $V(a_{1g})$ = 1.29 eV (this hybridization values are then scaled to 83\%), ligand crystal field = 0.693 eV. Slater integrals were reduced to 80\% of the Hartree-Fock values.}
}

@article{mouComparativeRamanScattering2024,
  title = {{Comparative {{Raman}} Scattering Study of Crystal Field Excitations in {{Co-based}} Quantum Magnets}},
  author = {Mou, Banasree S. and Zhang, Xinshu and Xiang, Li and Xu, Yuanyuan and Zhong, Ruidan and Cava, Robert J. and Zhou, Haidong and Jiang, Zhigang and Smirnov, Dmitry and Drichko, Natalia and Winter, Stephen M.},
  year = {2024},
  month = aug,
  journal = {Physical Review Materials},
  volume = {8},
  number = {8},
  pages = {084408},
  issn = {2475-9953},
  doi = {10.1103/PhysRevMaterials.8.084408},
  urldate = {2025-07-04},
  langid = {english}
}

@article{hussainExperimentalEvidenceCrystalfield2025a,
  title = {{Experimental Evidence of Crystal-Field, {{Zeeman-splitting}}, and Spin-Phonon Excitations in the Quantum Supersolid {{Na}}$_2${{BaCo}}({{PO}}$_4$)$_2$}},
  author = {Hussain, Ghulam and Zhang, Jianbo and Zhang, Man and Yadav, Lalit and Ding, Yang and Zheng, Changcheng and Haravifard, Sara and Wang, Xiawa},
  year = {2025},
  month = apr,
  journal = {Physical Review B},
  volume = {111},
  number = {15},
  pages = {155129},
  issn = {2469-9950, 2469-9969},
  doi = {10.1103/PhysRevB.111.155129},
  urldate = {2025-07-04},
  langid = {english}
}

@article{agrestiniProbingEff02018,
  title = {{Probing the {{J}}$_{\text{Eff}}$ = 0 Ground State and the {{Van Vleck}} Paramagnetism of the {{Ir}} $5^+$ Ions in Layered {{Sr}}$_2${{Co}}$_{0.5}${{Ir}}$_{0.5}${{O}}$_4$}},
  author = {Agrestini, S. and Kuo, C.-Y. and Chen, K. and Utsumi, Y. and Mikhailova, D. and Rogalev, A. and Wilhelm, F. and F{\"o}rster, T. and Matsumoto, A. and Takayama, T. and Takagi, H. and Haverkort, M. W. and Hu, Z. and Tjeng, L. H.},
  year = {2018},
  month = jun,
  journal = {Physical Review B},
  volume = {97},
  number = {21},
  pages = {214436},
  issn = {2469-9950, 2469-9969},
  doi = {10.1103/PhysRevB.97.214436},
  urldate = {2025-07-04},
  langid = {english}
}

@article{ferreira-carvalhoTrigonalDistortionKitaev2025,
  title = {{Trigonal Distortion in the {{Kitaev}} Candidate Honeycomb Magnet BaCo$_2$(AsO$_4$)$_2$}},
  author = {Ferreira-Carvalho, M. M. and R\"o\ss{}ler, S. and Chang, C. F. and Hu, Z. and Valvidares, S. M. and Gargiani, P. and Haverkort, M. W. and Mukharjee, Prashanta K. and Gegenwart, P. and Tsirlin, A. A. and Tjeng, L. H.},
  journal = {Physical Review B},
  volume = {112},
  issue = {12},
  pages = {125135},
  numpages = {10},
  year = {2025},
  month = sep,
  publisher = {American Physical Society},
  doi = {10.1103/2sdd-pyx1},
  url = {https://link.aps.org/doi/10.1103/2sdd-pyx1}
}

@article{popescuZeemanSplitKramers2025,
  title = {{Zeeman {{Split Kramers Doublets}} in {{Spin-Supersolid Candidate}} $\text{Na}_2\text{BaCo}(\text{PO}_4)_2$}},
  author = {Popescu, T. I. and Gora, N. and Demmel, F. and Xu, Z. and Zhong, R. and Williams, T. J. and Cava, R. J. and Xu, G. and Stock, C.},
  year = 2025,
  month = apr,
  journal = {Physical Review Letters},
  volume = {134},
  number = {13},
  pages = {136703},
  publisher = {American Physical Society},
  doi = {10.1103/PhysRevLett.134.136703}

}

@book{cowanTheoryAtomicStructure1981,
  title = {{The Theory of Atomic Structure and Spectra}},
  author = {Cowan, R. D.},
  year = {1981},
  month = sep,
  publisher = {University of California Press},
  isbn = {978-0-520-03821-9},
  langid = {english}
}

@article{bocquetElectronicStructureEarly1996,
  title = {{Electronic Structure of Early 3d-Transition-Metal Oxides by Analysis of the 2p Core-Level Photoemission Spectra}},
  author = {Bocquet, A. E. and Mizokawa, T. and Morikawa, K. and Fujimori, A. and Barman, S. R. and Maiti, K. and Sarma, D. D. and Tokura, Y. and Onoda, M.},
  date = {1996-01-15},
year = 1996,
month = jan,
  journal = {Physical Review B},
  volume = {53},
  number = {3},
  pages = {1161--1170},
  publisher = {American Physical Society},
  doi = {10.1103/PhysRevB.53.1161},
  url = {https://link.aps.org/doi/10.1103/PhysRevB.53.1161},


}

@article{hollmannLocalSymmetryMagnetic2010,
  title = {{Local Symmetry and Magnetic Anisotropy in Multiferroic ${\text{{{MnWO}}}}_{4}$ and Antiferromagnetic ${\text{{{CoWO}}}}_{4}$ Studied by Soft X-Ray Absorption Spectroscopy}},
  author = {Hollmann, N. and Hu, Z. and Willers, T. and Bohatý, L. and Becker, P. and Tanaka, A. and Hsieh, H. H. and Lin, H.-J. and Chen, C. T. and Tjeng, L. H.},
  date = {2010-11-23},
  year = 2010,
    month = nov,
  journal = {Physical Review B},
  volume = {82},
  number = {18},
  pages = {184429},
  publisher = {American Physical Society},
  doi = {10.1103/PhysRevB.82.184429},
  url = {https://link.aps.org/doi/10.1103/PhysRevB.82.184429},


}

@article{csiszarControllingOrbitalMoment2005b,
  title = {{Controlling {{Orbital Moment}} and {{Spin Orientation}} in {{CoO Layers}} by {{Strain}}}},
  author = {Csiszar, S. I. and Haverkort, M. W. and Hu, Z. and Tanaka, A. and Hsieh, H. H. and Lin, H.-J. and Chen, C. T. and Hibma, T. and Tjeng, L. H.},
  date = {2005-10-28},
  journal = {Physical Review Letters},
 year = 2005,
    month = oct,
  volume = {95},
  number = {18},
  pages = {187205},
  publisher = {American Physical Society},
  doi = {10.1103/PhysRevLett.95.187205},
  url = {https://link.aps.org/doi/10.1103/PhysRevLett.95.187205},

}

@article{vanveenendaalElectronicStructureCo2023a,
  title = {{Electronic Structure of {{Co}} $3d$ States in the {{Kitaev}} Material Candidate Honeycomb Cobaltate ${\text{{{Na}}}}_{3}{\text{{{Co}}}}_{2}{\text{{{SbO}}}}_{6}$ Probed with X-Ray Dichroism}},
  author = { van Veenendaal, M. and Poldi, E. H. T. and Veiga, L. S. I. and Bencok, P. and Fabbris, G. and Tartaglia, R. and McChesney, J. L. and Freeland, J. W. and Hemley, R. J. and Zheng, H. and Mitchell, J. F. and Yan, J.-Q. and Haskel, D.},
  date = {2023-06-28},
year = 2023,
    month = jun,
  journal = {Physical Review B},
  shortjournal = {Phys. Rev. B},
  volume = {107},
  number = {21},
  pages = {214443},
  publisher = {American Physical Society},
  doi = {10.1103/PhysRevB.107.214443},
  url = {https://link.aps.org/doi/10.1103/PhysRevB.107.214443},
}

@misc{carvalhoEdmondXe7bis2025,
author = {Carvalho, Miguel},
publisher = {Edmond},
title = {{Direct Evidence of a Near-Ideal Jeff = 1/2 Ground State in Triangular-Lattice Na2BaCo(PO4)2},https://doi.org/10.17617/3.XE7BIS},
year = {2025},
version = {V1},
doi = {10.17617/3.XE7BIS},
url = {https://doi.org/10.17617/3.XE7BIS}
}
\end{document}